\documentclass[aip,apl,twocolumn,superscriptaddress,groupedaddress]{revtex4}  
\usepackage{graphicx}  
\usepackage{dcolumn}   
\usepackage{bm}        
\usepackage{amssymb}   
\usepackage{upgreek}
\usepackage{subfigure}
\usepackage{float}
\usepackage{color}
\usepackage{soul}

\begin{document}


\title{Nested Trampoline Resonators for Optomechanics}

\author{M. J. Weaver} 
\email{mweaver@physics.ucsb.edu}
\affiliation{Department of Physics, University of California, Santa Barbara, California 93106, USA}
\author{B. Pepper}
\altaffiliation[Now at: ]{Jet Propulsion Laboratory, California Institute
of Technology, 4800 Oak Grove Drive, Pasadena,
California 91109, USA}
\affiliation{Department of Physics, University of California, Santa Barbara, California 93106, USA}
\author{F. Luna} 
\affiliation{Department of Physics, University of California, Santa Barbara, California 93106, USA}
\author{F. M. Buters}
\affiliation{Huygens-Kamerlingh Onnes Laboratorium, Universiteit Leiden, 2333 CA Leiden, The Netherlands}
\author{H. J. Eerkens}
\affiliation{Huygens-Kamerlingh Onnes Laboratorium, Universiteit Leiden, 2333 CA Leiden, The Netherlands}
\author{G. Welker}
\affiliation{Huygens-Kamerlingh Onnes Laboratorium, Universiteit Leiden, 2333 CA Leiden, The Netherlands}
\author{B. Perock}
\affiliation{Department of Physics, University of California, Santa Barbara, California 93106, USA}
\author{K. Heeck}
\affiliation{Huygens-Kamerlingh Onnes Laboratorium, Universiteit Leiden, 2333 CA Leiden, The Netherlands}
\author{S. de Man}
\affiliation{Huygens-Kamerlingh Onnes Laboratorium, Universiteit Leiden, 2333 CA Leiden, The Netherlands}
\author{D. Bouwmeester}
\affiliation{Department of Physics, University of California, Santa Barbara, California 93106, USA}
\affiliation{Huygens-Kamerlingh Onnes Laboratorium, Universiteit Leiden, 2333 CA Leiden, The Netherlands}

\date{\today}

\begin{abstract}
Two major challenges in the development of optomechanical devices are achieving a low mechanical and optical loss rate and vibration isolation from the environment. We address both issues by fabricating trampoline resonators made from low pressure chemical vapor deposition (LPCVD) Si$_3$N$_4$ with a distributed bragg reflector (DBR) mirror. We design a nested double resonator structure with 80 dB of mechanical isolation from the mounting surface at the inner resonator frequency, and we demonstrate up to 45 dB of isolation at lower frequencies in agreement with the design. We reliably fabricate devices with mechanical quality factors of around 400,000 at room temperature. In addition these devices were used to form optical cavities with finesse up to 181,000 $\pm$ 1,000. These promising parameters will enable experiments in the quantum regime with macroscopic mechanical resonators.
\end{abstract}

\maketitle


In recent years there has been tremendous growth in the field of optomechanics \cite{shortreview2012, review2014}. The interaction of light and mechanical motion has been used to demonstrate such phenomena as ground state cooling of a mechanical resonator \cite{cleland2010, painter2011, simmonds2011}, optomechanically induced transparency \cite{kippenberg2010, painterOMIT2011, simmondsOMIT2011}, and entanglement of a mechanical resonator with an electromagnetic field \cite{lehnert2013}. Another proposed application of optomechanics is testing the concept of quantum superpositions in large mass systems \cite{bouwmeester2003}. All of these experiments require low optical and mechanical loss rates. In this letter we will focus on our efforts to produce a large mass mechanical resonator with both high mechanical and optical quality factor, which can realistically be cooled to its ground state.

\begin{figure*}
\includegraphics[scale=0.55]{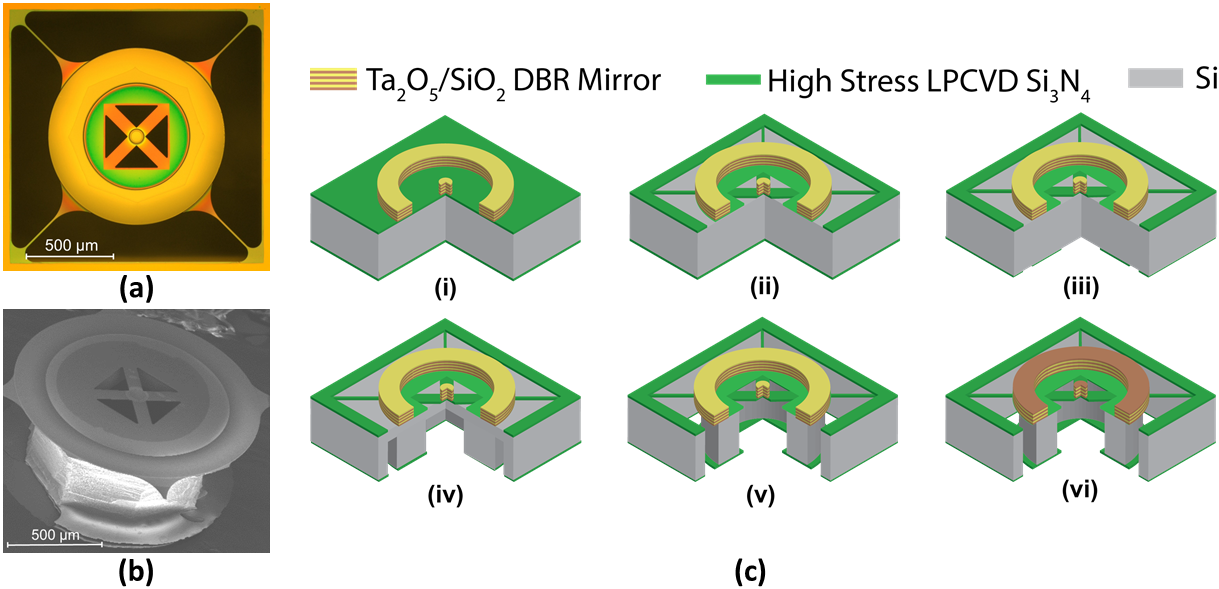}
\centering
\caption{Optical \textbf{(a)} and SEM \textbf{(b)} images of a nested trampoline resonator. The device was broken out of the chip to make the structure visible for \textbf{(b)}. Note the thin 10 $\upmu$m wide, 500 nm thick arms supporting the large 500 $\upmu$m thick silicon mass. A properly sized mirror layer was necessary to protect the nitride layer from sharp edges in the silicon and safely connect to the thin arms of the outer resonator. \textbf{(c)} A schematic overview of the fabrication process (not to scale). \textbf{(i)} The SiO$_2$/Ta$_2$O$_5$ DBR stack is etched via CHF$_3$ ICP etch. The front \textbf{(ii)} and back \textbf{(iii)} Si$_3$N$_4$ is etched by CF$_4$ plasma etch. \textbf{(iv)} Most of the Si is etched from the bottom using the Bosch process. \textbf{(v)} The remainder of the Si is etched via TMAH. \textbf{(vi)} A buffered HF dip cleans the devices and removes a protective SiO$_2$ layer. Only 6 layers of the SiO$_2$/Ta$_2$O$_5$ DBR stack are shown, and the shape of the outer resonator mass is approximated as a hollow cylinder for simplicity.}
\label{fig:fab}
\end{figure*}

There are several requirements for the devices to achieve this. The system must be sideband resolved for optical sideband cooling to the ground state \cite{girven2007, kippenberg2007}. A high mechanical quality factor is also necessary to generate a higher cooperativity and a lower mechanical mode temperature for the same cooling laser power. Furthermore, in the quantum regime, the quality factor sets the timescale of environmentally induced decoherence \cite{zurek2003}, which is crucial for proposed future experiments. Therefore, it is important to eliminate mechanical and optical loss sources.

One major source of loss in mechanical systems is clamping loss, which is coupling to external mechanical modes \cite{aspelmeyer2011, parpia2011, regal2012}. As we will show, this is a critical source of loss for Si$_3$N$_4$ trampoline resonators. Several methods of mechanically isolating a device from clamping loss have been demonstrated including phononic crystals \cite{painterPC2011, regalPC2011} and low frequency mechanical resonators \cite{norte2014, bonaldi2012, liu2014, haringx1947, LIGO2010}. Due to the large size of phononic crystals at the frequency of our devices (about 250 kHz), we have selected to surround our devices with a low frequency outer resonator. We significantly improve on the design of similar devices using silicon optomechanical resonators \cite{bonaldi2015} by using a lower frequency outer resonator and silicon nitride with weaker spring constant. Weaker spring constants lead to higher optomechanical coupling, a requirement for our future experiments. The outer resonator acts as a mechanical second order low pass filter with the following mechanical transfer function \cite{thorntonmarion}:

\begin{equation}
T(\omega) = \frac{\omega_o^4}{(\omega_o^2-\omega^2)^2+\gamma_o^2\omega^2}
\label{eq:transfer}
\end{equation}
$\omega$ is the frequency of vibration, $\omega_o$ is the frequency of the outer resonator and $\gamma_o$ is the mechanical loss rate of the outer resonator. Choosing an outer resonator frequency of 2.5 kHz and an inner resonator frequency of 250 kHz leads to approximately 80 dB of isolation of the inner resonator. This isolation is independent of $\gamma_o$ \footnote{If $\omega >> \omega_o$, the transfer function is well approximated as $T(\omega) = \omega_o^4/\omega^4$, which falls off at 40 dB per decade and is independent of the outer resonator quality factor.}. The nested trampoline resonator scheme promises both a high mechanical quality factor independent of mounting and mechanical isolation from the environment.

Our optomechanical system is a 5 cm long Fabry-P\'{e}rot cavity consisting of a large distributed bragg reflector (DBR) mirror deposited on a SiO$_2$ curved surface and a nested trampoline resonator device. The nested trampoline resonator has a small DBR mirror (80 $\upmu$m in diameter) mounted on four Si$_3$N$_4$ arms, surrounded by a large silicon mass held in place by four more Si$_3$N$_4$ arms (See Figure \ref{fig:fab}). Previously, we have fabricated single resonator devices with plasma enhanced chemical vapor deposition (PECVD) low stress nitride \cite{bouwmeester2011}. In this letter, we use high stress low pressure chemical vapor deposition (LPCVD) Si$_3$N$_4$, because it generally has higher frequency and lower intrinsic loss \cite{kotthaus2010}. The stress is typically around 1 GPa for LPCVD Si$_3$N$_4$ \cite{springer}, but comparisons between Finite Element Analysis models and the observed frequencies of fabricated devices indicate that the stress is probably closer to 850 MPa in this case.


Devices are fabricated starting with a superpolished 500 micron thick silicon wafer. Either 300 or 500 nm of high stress LPCVD Si$_3$N$_4$ is deposited on both sides of the wafer, and a commercially procured SiO$_2$/Ta$_2$O$_5$ DBR is deposited on top. The DBR is etched into a small mirror on the inner resonator and a protective mirror layer on the outer resonator using a CHF$_3$ inductively coupled plasma (ICP) etch. Next, the Si$_3$N$_4$ arms of the devices are patterned with a CF$_4$ etch. A window is also opened on the back side Si$_3$N$_4$ with a CF$_4$ etch. Approximately 400 microns of silicon under the Si$_3$N$_4$ arms are removed from the back using the Bosch deep reactive ion etch process. A large silicon mass is left in place between the inner and outer arms of the device. The devices are then released with a tetramethylammonium hydroxide (TMAH) etch. Finally a buffered HF etch removes the top protective buffer layer of SiO$_2$ without damaging the underlying Ta$_2$O$_5$ layer. Figure \ref{fig:fab} shows a schematic summary of the fabrication process.

Devices are characterized using a 1064 nm NdYAG laser. To measure mechanical motion, the Fabry-P\'{e}rot cavity is first intentionally misaligned to a finesse of around 100 to avoid any optomechanical effects. The cavity is then locked to the laser frequency at the inflection point of a Fabry-P\'{e}rot fringe using a piezoelectric actuator moving the position of the large mirror. Quality factors are taken from Lorentzian fits to the power spectral density of the Brownian motion of the devices. Finesse is measured by optical ringdown \cite{bouwmeester2011}.



\begin{figure}
\includegraphics[scale=0.7]{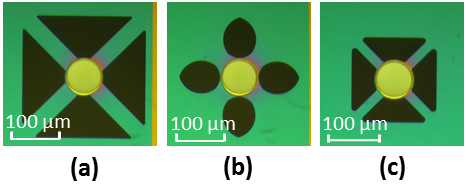}
\caption{Optical microscope images of three single resonator devices. A number of different geometries were fabricated with different arm length, arm width and fillet size.}
\label{fig:singledevs}
\end{figure}

As an initial step, a series of single trampoline resonators with 60 $\upmu$m diameter mirrors and varying geometries were fabricated and the mechanical quality factors measured \cite{bouwmeesterPRA2015, bouwmeester2015}. Three of the devices are pictured in Figure \ref{fig:singledevs}. We observed no significant geometric trends in quality factor. However, we found that remounting the same sample can change the quality factor of the devices by more than a factor of 10. Table \ref{tab:singleqs} shows the quality factors for the devices on one chip mounted three separate times. It is clear that mounting drastically affects the quality factor; we attribute this to a change in the clamping loss, because we observe mechanical modes in the system around the resonance frequency that change in number, frequency and power with mounting. Clamping loss can be modeled as a coupling to these external mechanical modes \cite{wilsonrae2008, aspelmeyer2011}.


We now turn to the nested trampoline resonators (see Figure \ref{fig:fab}.) The outer resonator acts as a low pass filter, providing 40 dB of isolation for every decade of frequency difference between the inner and the outer resonator (see Equation \ref{eq:transfer}.) To test the mechanical isolation we performed a vibration transmission experiment. We attached a ring piezo to the sample mount with springs and applied a sinusoidal signal of varying frequency to the piezo. We measured the motion of the chip using a Michelson interferometer and the motion of the inner mirror using a low finesse Fabry-P\'{e}rot cavity as described above. The ratio of these two signals is the mechanical transmission from the chip mounting to the inner mirror. 

\begin{figure}[t]
\includegraphics[scale=0.46]{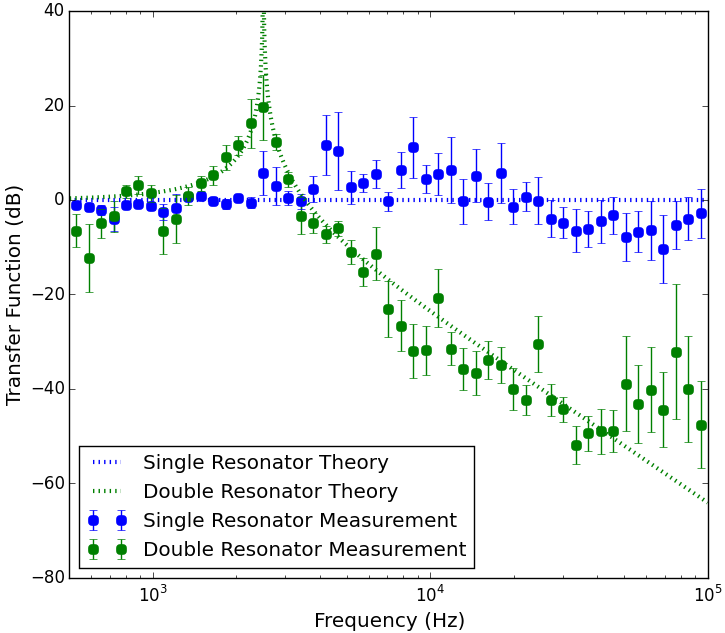}
\caption{Transfer function of a single and nested resonator. A sample mount with a single (blue) and nested (green) resonator was mechanically driven at a range of frequencies. The motion of the outer chip and the inner mirror were measured to get the mechanical transfer function. The height at DC frequencies is adjusted to zero. This plot demonstrates that the nested resonator scheme provides mechanical isolation as predicted by Equation \ref{eq:transfer}.}
\label{fig:transfer}
\end{figure}

This challenging experiment required eight orders of magnitude to be measured in the same frequency scan. Because of insufficient laser scanning range, the Michelson interferometer was uncalibrated and the DC response was used for calibration. Due to the requirement for a single scan, measurement averaging time was limited by drift in the interferometer. The mechanical response of the piezo also dropped off significantly after 100 kHz, so it was not possible to measure the mechanical transmission at the frequency of the inner resonator. Figure \ref{fig:transfer} shows the transmission for both a single and a nested resonator. The data are binned for clarity, with the error bars reflecting variations within each bin. The experimental data follow the trend predicted by Equation \ref{eq:transfer} quite well. The theory curve is not a fit; $\omega_o$ and $\gamma_o$ were determined through independent measurements. The deviations at high frequency are likely due to insufficient signal to noise ratio. The results clearly indicate that the outer resonator provides approximately 40 dB per decade of mechanical isolation. We can only measure a maximum of 45 dB of isolation, but we would expect 80 dB of isolation if we continued the measurement up to the inner resonator frequency.
\begin{table}
\begin{tabular}{c | c | c | c }
\hline \hline
Mounting & Device a & Device b & Device c \\
\hline
1 & 425,000 $\pm$ 32,000 & 80,000 $\pm$ 4,000 & 33,000 $\pm$ 2,000  \\
2 & 38,000 $\pm$ 2,000 & 5,000 $\pm$ 1,000 & 40,000 $\pm$ 2,000  \\
3 & 264,000 $\pm$ 21,000 & 16,000 $\pm$ 1,000 & 113,000 $\pm$ 8,000  \\
\hline \hline
\end{tabular}
\caption{This table shows the quality factors for the three devices pictured in Figure \ref{fig:singledevs} with three different mountings. The importance of clamping loss is evident from the changes in quality factor of more than a factor of ten based on the mounting.}
\label{tab:singleqs}
\end{table}

\begin{figure}[H]
\includegraphics[scale=0.42]{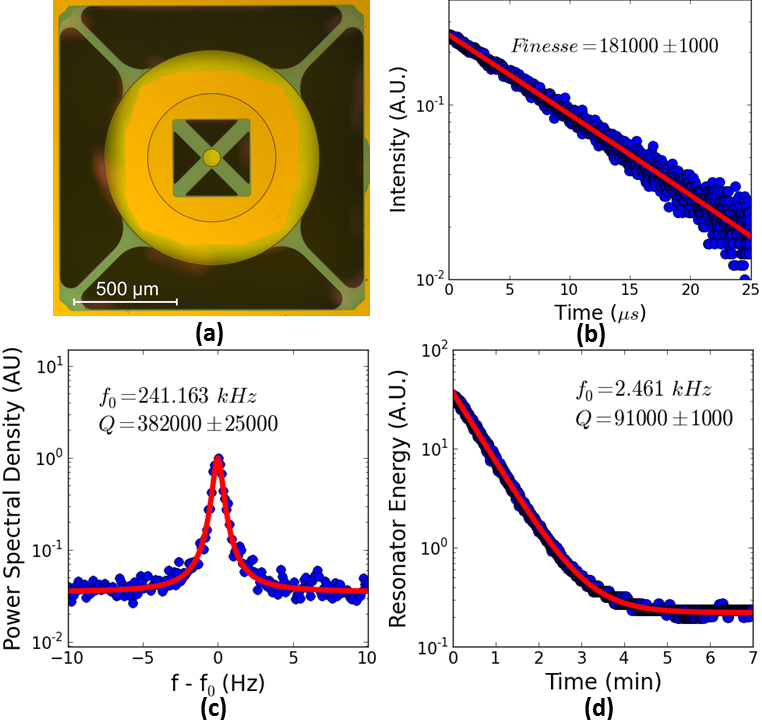}
\caption{Measurements of a nested trampoline resonator (the same device as for Figure \ref{fig:transfer}.) (a) Optical microscope image. (b) Optical ringdown to measure the cavity finesse. (c) Lorentzian fit to thermal motion of the inner resonator to measure quality factor.  (d) Mechanical ringdown of the outer resonator to measure quality factor taken using a lock-in amplifier.}
\label{fig:dev2}
\end{figure}

We also tested the mounting dependence of the quality factor. The results of remounting a single nested resonator five times are shown in Table \ref{tab:measurements}. The quality factor of the outer resonator changes drastically between the mountings, indicating that the mechanical clamping loss is changing. However, the inner resonator only demonstrates changes in quality factor on the order of 10$\%$. The relatively small variation in quality factor of the inner resonator and the absense of extra mechanical peaks around the resonance frequency indicate that the clamping loss of the device has largely been eliminated. Indeed, all nested resonators fabricated without any obvious physical defects had quality factors between 300,000 and 500,000. The highest quality factor achieved was 481,000 $\pm$ 12,000, an order of magnitude larger than for comparable silicon devices at room temperature \cite{bonaldi2015}. Typical quality factor measurements for an inner and outer resonator are shown in Figure \ref{fig:dev2}.

\begin{table}
\begin{tabular}{c | c | c }
\hline \hline
Mounting & Inner Resonator Q & Outer Resonator Q \\
\hline
1 & 418,000 $\pm$ 11,000 & 700,000 $\pm$ 100,000 \\
2 & 427,000 $\pm$ 10,000 & 690,000 $\pm$ 100,000 \\
3 & 481,000 $\pm$ 12,000 & 70,000 $\pm$ 20,000 \\
4 & 462,000 $\pm$ 14,000 & 240,000 $\pm$ 40,000 \\
5 & 457,000 $\pm$ 13,000 & 220,000 $\pm$ 40,000 \\
\hline \hline
\end{tabular}
\caption{This table shows the quality factors of a nested trampoline resonator remounted five different times. The outer resonator quality factor (measured via ringdown) has large variation between the mountings while the inner resonator quality factor (measured via a fit to thermal motion) has only small variation between the mountings.}
\label{tab:measurements}
\end{table}

One concern for experiments with this system is the thermal motion of the outer resonator (10-100 pm rms at room temperature). Because of the narrow linewidth of the cavity, the optical response to such a large motion is nonlinear. However, the frequency of the outer resonator is low enough that a PID controller can lock a laser to the cavity, tracking the motion and removing any nonlinear effects. In addition, if the laser is locked with a slight negative detuning from the cavity resonance, the outer resonator can be optomechanically cooled, even without being sideband resolved \cite{review2014}. Thus, the motion of the outer resonator does not prevent experiments using the inner resonator.

Another concern is maintaining the high quality of the DBR mirror layer through the fabrication process. Reducing the optical loss rate is critical to developing a system that allows quantum optical manipulation of mechanical motion. One way to reduce the optical loss rate is through superpolishing the wafer surfaces before deposition of the DBR, to reduce scattering. The addition of this step, as well as the selection of very highly reflective DBR coatings enable us to achieve a Fabry-P\'{e}rot cavity with finesse 181,000 $\pm$ 1,000, (optical linewidth 17 kHz) the highest finesse reported in an optomechanical Fabry-P\'{e}rot system. The ringdown measurement is shown in Figure \ref{fig:dev2}. All of the nested resonators measured have finesse greater than 160,000, indicating that the nested trampoline fabrication process is completely compatible with maintaining highly reflective mirror surfaces.


Improvements in finesse and mechanics will enable new experiments with trampoline resonators. Our system (using the device in Figure \ref{fig:dev2}) is fourteen times sideband resolved, which is more than sufficient for experiments such as quantum nondemolition measurements \cite{jacobs2008}. The elimination of the clamping loss will enable another systematic study of the geometry like the one attempted with single resonators. Many mechanical devices using Si$_3$N$_4$ without a DBR have much higher quality factor \cite{harris2008, regalPC2011, norte2014}. Varying the design of the inner resonator could allow reduction of mirror-nitride loss and fabrication of devices with even higher quality factors.

The improvements in mechanical isolation should also enable optomechanical cooling to the ground state. The devices are shielded from environmental mechanical noise, which previously could obscure motion at the quantum level. The fQ product of 1.1x10$^{11}$ Hz (for the device from Table II) is also high enough for cooling to the ground state from 4 K, potentially alleviating the need for a dilution refrigerator. Our sideband resolution yields a theoretical minimum of 3 x 10$^{-4}$ phonons from optical cooling if there is no heating of the system \cite{girven2007}.One concern is the thermal conductivity of our design, because at 4 K the thermal conductivity of Si$_3$N$_4$ drops to about 10$^{-2}$ W/mK \cite{pekola1998,bourgeois2015}. The heat conduction is limited by the arms of the outer resonator, which are five to fifteen times narrower than the arms of the inner resonator. We have previously thermalized single 
 \noindent
resonators to 100 mK temperature, so thermalizing a double resonator sample to 4K, even with the narrower arms, should not be a problem.  

We have demonstrated that we can consistently fabricate nested trampoline devices with both high quality factor and high finesse. We design the devices to have 80 dB of mechanical isolation from the environment at the inner resonator frequnecy, and we observe greater than 45 dB of mechanical isolation at lower frequencies and the elimination of clampling losses. These high quality parameters will pave the way to fabrication of even better devices and measurements at low temperatures.

\section{Acknowledgments}

The authors would like to thank H. van der Meer for
technical assistance and support and P. Sonin for helpful discussion about the design of single resonators. This work is supported by National Science Foundation
Grant No. PHY-1212483. This work is also part of the research program of the
Foundation for Fundamental Research on Matter (FOM) and
of the NWO VICI research program, which are both part of
the Netherlands Organisation for Scientific Research (NWO).

\bibliography{NestedTrampolinesAPLRevised}

\end{document}